\documentclass[epj]{svjour}

\usepackage{graphicx}
\def\bea{\begin{eqnarray}}
\def\eea{\end{eqnarray}}

\begin{document}
\title{Cellular solid behaviour of liquid crystal colloids \\
2. Mechanical properties}
\author{V.J. Anderson \and E.M. Terentjev}
\institute{ Cavendish Laboratory, University of Cambridge,
Madingley Road, Cambridge CB3 0HE, U.K. }
\date{\today}

\abstract{ This paper presents the results of a rheological study
of thermotropic nematic colloids aggregated into cellular
structures. Small sterically stabilised PMMA particles dispersed
in a liquid crystal matrix densely pack on cell interfaces, but
reversibly mix with the matrix when the system is heated above
$T_{\rm ni}$. We obtain a remarkably high elastic modulus, $G'
\geq 10^5 \, \hbox{Pa}$, which is a nearly linear function of
particle concentration. A characteristic yield stress is required
to disrupt the continuity of cellular structure and liquify the
response. The colloid aggregation in a ``poor nematic'' MBBA has
the same cellular morphology as in the ``good nematic'' 5CB, but
the elastic strength is at least an order of magnitude lower.
These findings are supported by theoretical arguments based on the
high surface tension interfaces of a foam-like cellular system,
taking into account the local melting of nematic liquid and the
depletion locking of packed particles on interfaces.
\PACS{ {61.30.-v}{Liquid crystals.} \and
       {82.70.-y}{Disperse systems.} \and
       {81.40.Jj}{Elasticity and anelasticity, stress-strain relations.}
     }
} %end of abstract
\authorrunning{V.J. Anderson {\it et al.}}
\titlerunning{Cellular solid liquid crystal colloids: 2.
Mechanical properties}
 \maketitle
\section{Introduction}  \label{intro}
In the preceding paper \cite{no1} we have described the structure
and aggregation mechanisms of thermotropic nematic colloid with a
new cellular morphology. We argued that, in some circumstances,
the continuous phase separation takes place, driven by the mean
field energy of orientational nematic order in the matrix. This
mechanism, and the resulting cellular structure with colloid
particles densely packed on thin interfaces, is different from a
number of previous observations
\cite{cladis,roux,poulin98,martin}. In those cases particles
always aggregated into dense 3-dimensional flocs, often supported
by line or point disclinations.

We showed that the behaviour of a liquid crystal colloid is
determined by the dimensionless parameter $WR/K$, where $W$ is the
nematic anchoring energy on the particle surface, $K$ the Frank
elastic constant\footnote{For typical thermotropic nematics and
homeotropic director anchoring used in our work, these material
constants take the values $W \sim 10^{-6} \hbox{J/m}^2$ and $K\sim
10^{-11} \hbox{J/m}$, e.g. \cite{degen}.} and $R$ the particle
radius. A spherical colloid particle in a nematic liquid crystal
creates a topological defect in the director field, if $WR/K \gg
1$. When this parameter is small (which is our case, with PMMA
particles of $R\sim 150 \, \hbox{nm}$), the particle introduces
only a small perturbation into the nematic matrix. In this case
the long range interaction forces between particles are
sufficiently weak so that the aggregation times are long compared
with the alternative mechanism of continuous phase separation.
Direct observation in the bulk of our samples with the confocal
microscopy technique have confirmed the open-cell morphology and
the estimate for the cell size of the structure, $\lambda \sim
(nR)/\Phi$, where $nR$ is the wall thickness and $\Phi$ the
average particle concentration, (experimentally we find $n \sim
20-30$, \cite{no1}).
\begin{figure}[b]
\centerline{\resizebox{0.47\textwidth}{!}{\includegraphics{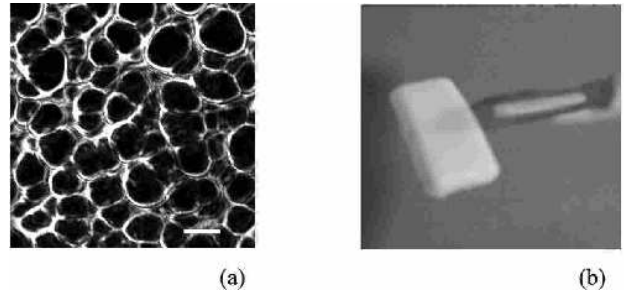}
} }
 \caption{(a) Reflection-mode confocal image of cellular structure
in the 5CB nematic colloid at $\Phi=10\%$ \cite{no1}. The white
bar shows the length scale of 20$\mu$m. \ (b) The photo of an
aggregated 10\%-5CB colloid, showing its white scattering
appearance and the evident rigidity. } \label{fig1}
\end{figure}
Recently Tanaka {\it et al.} \cite{tanaka} studied phase
separation in mixture of a 5CB nematic and a surfactant, which
forms very small micelles. They reported some provocative results,
also quite different from ours. We could speculate that, in their
case, the colloid parameter $WR/K$ is so small that the mechanism
of phase separation described in \cite{no1} is not valid, indeed,
nano-micelles might be too small to apply any mean-field
description of a nematic order.

In any case, regardless of which structure the aggregating liquid
crystal colloid adopts, the phase equilibrium and kinetics are
altered in non-trivial ways due to the underlying frustrated
liquid crystalline order. The cellular structure we find in our
systems is, clearly, a metastable state. One of the main questions
we address here is the change in rheological behaviour due to
locally highly concentrated forces in the aggregated networks,
creating high barriers to deformation. One may expect a glass-like
freezing of motion and the resulting prolonged stability of the
phases.

In contrast to the network supported by topological defects
(disclinations), as in the recent study \cite{martin} of bigger
aggregates and somewhat weak static modulus $G' \sim 0.01 \,
\hbox{Pa}$, the cellular solids in our work possess a storage
modulus of $G' \geq 10^5 \, \hbox{Pa}$. This solidification was
first reported in \cite{wilson} for a 5CB-PMMA colloid with
particles of $R \approx 250$~nm. In this paper we present further
results on the mechanical properties of nematic colloid, with
smaller particles and fast quenching rates, showing the storage
modulus approaching an even higher values, $10^6 \, \hbox{Pa}$,
and examining the yield and ageing properties of this phenomenon.

\section{Experimental}
The preparation procedures were described in some detail in the
companion paper \cite{no1}. A typical thermotropic nematic liquid
crystal 5CB (from Aldrich) has been mixed with monodisperse PMMA
particles of radius $R=150 \, \hbox{nm}$, sterically stabilised by
chemically grafted poly-$12$-hydroxy stearic acid chains. Such
short radially grafted polymers provide a homeotropic director
anchoring. In all experiments we use the relatively small particle
volume fraction, not more than $\Phi=15\%$, determined by weight.
Good mixing of particles was achieved above the nematic-isotropic
transition and the samples were stored at $T\sim 45$C (above
$T_{\rm ni}=35.8$C in pure 5CB) in a tumbling device to ensure
that mixtures are homogeneous before any experiment is started.

An alternative colloid system has been prepared with another
famous thermotropic nematic MBBA (Aldrich). We use this system for
comparison, as an example of a ``poor'' nematic order in the
supporting matrix (the MBBA is partially hydrolysed in ambient
conditions, with effective $T_{\rm ni}=37.3$C in pure MBBA), see
\cite{no1} for detail.

 The mechanical properties have been studied on a Dynamic Stress
Rheometer, from Rheometrics Ltd, in the parallel plate set-up. A
sinusoidal stress of low frequency is applied to the sample and
the resulting strain is measured and interpreted by the software.
Unfortunately, this device did not have an option to impose an
oscillating strain of small constant amplitude. This prevented us
from studying in detail the sample rheology during the
nematic-isotropic transition and structure formation -- the change
in the modulus is so great (between $\sim 0.01$ and $10^5$~Pa)
that the constant-stress mode of the rheometer could not cope and
the settings had to be adjusted below $T_{\rm ni}$.

The rheometer was equipped with the purpose-made environmental
stage allowing the rapid cooling of the sample (at a rate $35^{\rm
o}$/min). The experimental procedure was as following:\\
 1) The sample was heated to $\sim 50$C in an
ultrasonic bath (to ensure homogeneous particle mixing) and then
transferred, via a pre-heated syringe, to the pre-heated rheometer
plate. The upper plate was then lowered slowly, to the gap $\sim
0.5$~mm. All through this manipulation, the temperature of the
sample was remaining above its $T_{\rm ni}$.
\\
 2) The environmental chamber was then closed and the temperature
stabilised at $50$C to return the sample to its isotropic mixed
state. The oscillating movement of plates was started at this
moment, with a low constant value of stress $\sim 0.3$~Pa since
the isotropic suspension is a liquid with low viscosity, and a
constant frequency of $1$~Hz. Samples were allowed to equilibrate
in this regime for at least $10$~min.
\\
 3) The rheometer run was then stopped and the
temperature quickly dropped to the working level, every time 15
degrees below $T_{\rm ni}$ (e.g. to $20$C for $\Phi=5\%$ and $18$C
for $\Phi=15\%$ in 5CB). The rheometer was restarted at exactly
100~s after the working temperature was reached. A higher value of
oscillating stress amplitude, $\sim 3-30$~Pa, was required to
acquire reliable data on the resulting rigid cellular solid.
\\
 4) To study the yield stress behaviour,
the same procedure of sample loading and quenching was followed.
The rheometer was then set to perform a low frequency stress-sweep
test, measuring the response of the system on increasing shear
stress -- thus reaching the point of breaking and yield of the
cellular structure.

\section{Results and discussion}
 The first observation that one makes on cooling the
homogeneously mixed colloid suspension below the nematic-isotropic
transition is the rapid optical change. Above $T_{\rm ni}$ the
suspension with a low particle concentration is relatively
transparent. Below the clearing point the colloid becomes
completely opaque, white in the case of 5CB and marginally yellow
for MBBA. Clearly, a strong multiple scattering of light takes
place in the material -- significantly greater than in a pure
nematic liquid crystal, which is turbid due to the director
thermal fluctuations \cite{degen}. Such an opaqueness is usually a
signature of randomly quenched disorder in the nematic director
field and is found, for instance, in nematic silica-gels
\cite{bellini} and polydomain nematic elastomers \cite{rev}. Our
study of structure of the aggregated nematic colloid below $T_{\rm
ni}$ in the companion paper \cite{no1} confirms this view: the
director is anchored on the walls of the cellular structure, see
fig.~\ref{fig1}(a).

Secondly, we find the remarkable rigidity of the resulting
material. Even before the detailed rheological study, one notices
that the opaque aggregated colloid is solid. It evidently supports
its own weight, fig.~\ref{fig1}(b), one can cut it with a knife,
imprint shapes and handle it quite robustly. An important
unresolved question is of the ageing of such a system. In general,
the prepared shape of the aggregated cellular solid is preserved
for a long time, if the temperature is kept constantly deep below
$T_{\rm ni}$. However, we observe that some amount of liquid
(apparently, the pure liquid crystal) leaks from the samples over
a period of days and weeks, the effect being much stronger in MBBA
colloids than in 5CB ones, and at smaller particle
concentrations.\footnote{In fact, there is quite a difference
between the 5CB and MBBA colloids. The MBBA-material is much less
robust, breaks, leaks liquid and does not have such a long
``shelf-life'' as the 5CB one.} In spite of this, the integrity of
the sample solid shape seemed to be preserved. In this paper we do
not discuss effects of ageing and leave it for another, more
detailed research work.

A third general observation, made both during the uncontrolled
cooling of the colloid mixture on the lab bench and in the
environmental chamber of the rheometer, is that the rate of
cooling has a strong effect on the resulting rigidity. Fast
quenching, at least in the initial period after reaching the
working temperature, makes the material much more rigid and the
cellular structure more regular. The latter fact is obtained by
comparing the confocal microscope images of cellular morphology of
the same samples cooled on the bench and those taken from the
rheometer after a fast quenching and parallel-plate rheometer run.

Finally, turning to the detailed study of the mechanical
properties of aggregated cellular solids, we find that another
crucial effect is the actual value of the working temperature: the
storage modulus $G'$ depends on it very strongly. This introduces
a slight ambiguity in comparison of the data between colloids of
different particle concentration, since they have their $T_{\rm
ni}$ different, decreasing with the average volume fraction of a
colloid mixture $\Phi$ \cite{no1}. We chose to work at a fixed
difference below $T_{\rm ni}$, but it is not clear that the
resulting difference in absolute values of working $T$ does not
introduce a shift in $G'$. Due to this problem, one cannot be
certain that the plots of $G'$ against colloid concentration,
figs.~\ref{modulus}(b) and \ref{modulus2}(b), are completely
consistent and one should interpret them with a degree of caution.

\subsection{Storage modulus}
%%%%
\begin{figure} %[b]
\centerline{\resizebox{0.45\textwidth}{!}{
\includegraphics{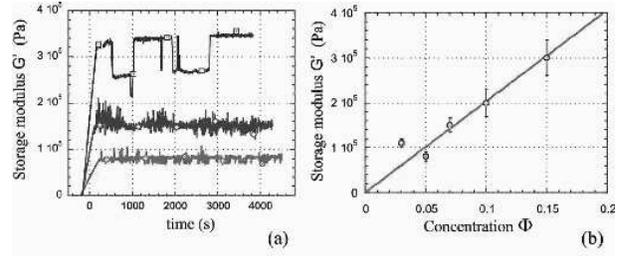}
}}
 \caption{The evolution with time of the
storage modulus $G'$ at constant frequency $\omega =1$~Hz for the
5CB colloid at $\Delta T \approx 15^{\rm o}$ below $T_{\rm ni}$,
for particle concentrations $\Phi=5\%, \ 7\%$ and $15\%$, graph
(a), curves with increasing saturation. Graph (b) shows the
dependence of the saturation value $G'$ on the particle
concentration. A linear fit gives, $G' \sim \left(2 \times 10^6
\right) \, \Phi  \, \hbox{Pa}$. } \label{modulus}
\end{figure}
%%%%
%%%%
\begin{figure} %[b]
\centerline{\resizebox{0.45\textwidth}{!}{
\includegraphics{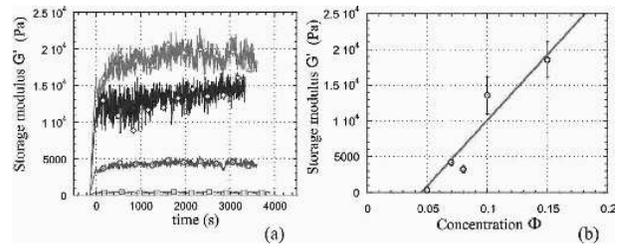} }}
 \caption{The evolution with time of the
storage modulus $G'$ at constant frequency $\omega =1$~Hz for the
MBBA colloid at $\Delta T \approx 15^{\rm o}$ below $T_{\rm ni}$,
for particle concentrations $\Phi=5\%, \ 7\%, \ 10\%$ and $15\%$,
graph (a), curves with increasing saturation. Graph (b) shows the
dependence of the saturation value $G'$ on the particle
concentration. An attempted linear fit gives $G' \sim \left(1.8
\times 10^5 \right) \, [\Phi-0.045] \, \hbox{Pa}$. }
\label{modulus2}
\end{figure}
%%%%
After following the same thermal history for all the samples, that
is a rapid quench to $15^{\rm o}$ below their respective $T_{\rm
ni}$ and a 100~s delay after this temperature was first reached,
the complex modulus $G(\omega)$ of the aggregated cellular solids
has been measured as a function of time. We are particularly
interested in the storage modulus which, at low frequency such as
$\omega=1$~Hz, is a representative feature of the equilibrium
elastic response. Of course, above $T_{\rm ni}$, for a
homogeneously mixed low-concentration colloid suspension, this
storage modulus is essentially zero.

The resulting values of storage modulus $G'$ are remarkably high
for all materials and compositions studied -- this solidification
of a nematic colloid is the main result of this paper, illustrated
in figs.~\ref{modulus} and \ref{modulus2}. The accurate
measurement of this modulus meets some unavoidable experimental
difficulties: for a large $G'$, the stress applied to the
rheometer plates has to be sufficiently high to produce a
measurable strain. However, too high a stress results in the
slipping of the plates, which become lubricated by the leaking
nematic liquid, or the complete breaking of the cellular structure
(the yield effect, discussed below). The compromise between these
two trends leads to rather high experimental noise, especially in
systems with higher particle concentration where the shear strain
tends to be lower for an allowed range of applied stress.

The results for $G'$ in 5CB and MBBA aggregated colloids show
small variation of the modulus with time, at least during the
first several minutes or hours. There is also a marked difference
in values between the 5CB colloid and the ``poor'' nematic
material based on partially hydrolysed MBBA. The elastic modulus
is more than an order of magnitude lower in the MBBA colloid.

Formation of a rigid cellular solid proven, the final level of
$G'$ is related to the initial particle concentration.
Figures~\ref{modulus}(b) and \ref{modulus2}(b) give such a
dependence. The data points are somewhat scattered and are
affected by a high noise, or experimental uncertainty. However,
the linear dependence on concentration is a plausible conclusion
for both 5CB and MBBA colloid aggregates. A striking difference
between the two systems, apart from the magnitude of $G'$, is the
apparent low-concentration threshold at $\Phi \sim 4.5\%$ for the
cellular aggregation of the MBBA colloid. This observation in the
``poor'' nematic model system is consistent with the phase diagram
in fig.~9 of the companion paper \cite{no1}. There, examining the
case of smaller strength of nematic field, one finds a noticeable
region of miscibility at low concentrations, below $T_{\rm ni}$.
In contrast, for the Landau parameters characteristic of a
``strongly nematic'' 5CB, this region is too narrow to be
practically observable -- accordingly, the $G'(\Phi)$ dependence
for 5CB suggests that it aggregates from nearly zero $\Phi$.

At this point one may remark that the rheometric data obtained at
a slow cooling rate \cite{wilson} have shown a much lower values
of storage modulus $G'$. Although there is a difference in colloid
particle sizes (the authors of \cite{wilson} used $R \approx
250$~nm, which should affect the aggregation kinetics and
morphology), we also find that the rate of system quenching and
the time the aggregating colloid spends at rest at the working
temperature have a strong effect on the resulting strength of
cellular structure.

\subsection{Stability and Yield stress}

The aggregated cellular structures of nematic colloids were very
stable over the period of about an hour, as confirmed both by
visual observation, constancy of confocal images and the little
variation of measured $G'$ with time. Over longer periods, one
observes a ``leaking'' of pure liquid crystal from the rather
rigid and completely opaque cellular aggregate. We did not study
this effect in detail, but one may speculate that the open-cell
morphology allows the nematic liquid to gradually find its way out
of the over-constrained cellular environment.

However, from the moment of formation one observes a systematic
step-like variation in the modulus, especially visible for 15\%
5CB system in fig.~\ref{modulus}(a). In this particular material
the $G'$ is so high that we had to apply a relatively high
constant stress to obtain any measurable deformation response. One
might argue that the the steps between two distinct values of 2.6
and 3.5$\times 10^5$~Pa are due to a partial breaking and
reforming of the cellular structure.\footnote{An alternative
possibility of rheometer plate slipping was also frequently
observed, but this always resulted in a rapid monotonic decline of
measured $G'$ to nearly zero -- not the repeated steps as in
fig.~\ref{modulus}(a).}

%%%%
\begin{figure} %[h]
\centerline{\resizebox{0.45\textwidth}{!}{
\includegraphics{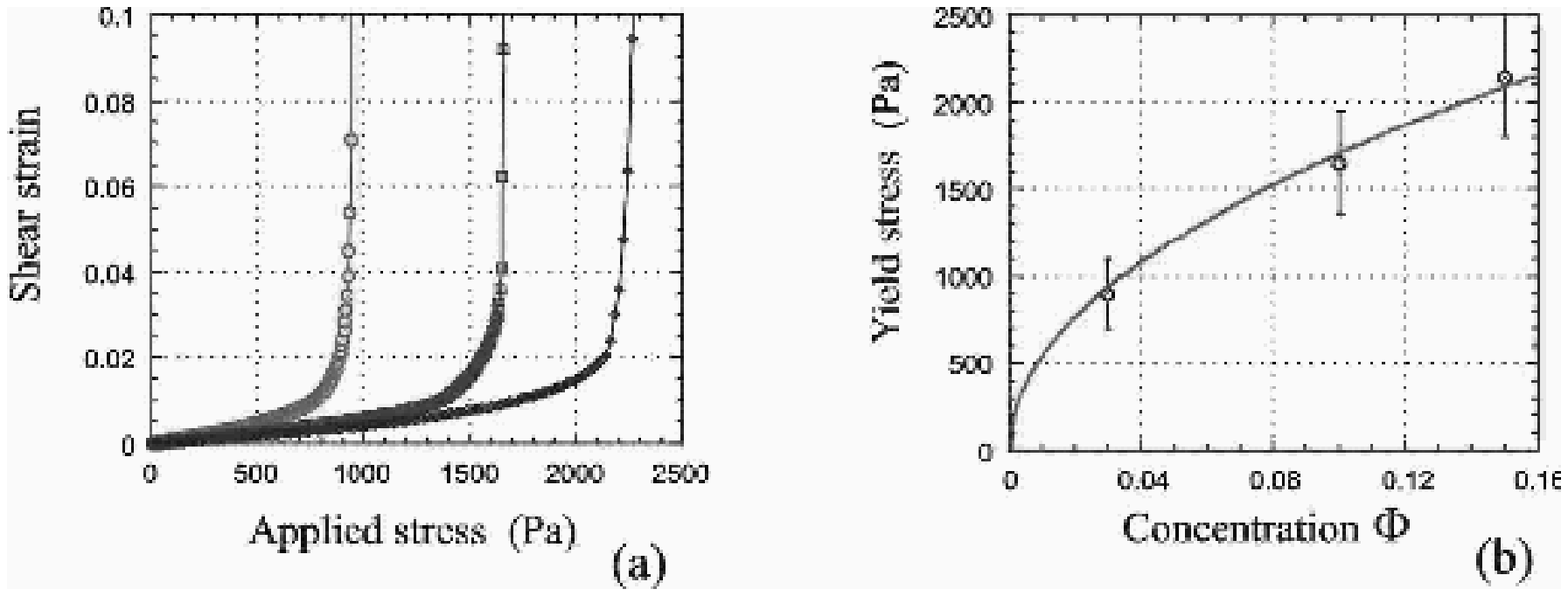}}  }
 \caption{The yield stress effect: (a) Strain-stress curves at a frequency
$\omega =1$~Hz for the 5CB colloid at $\Delta T \approx 15^{\rm
o}$ below $T_{\rm ni}$, for particle concentrations $\Phi=3\%, \,
10\%$ and $15\%$.  \ (b) The yield stress vs particle
concentration; solid line shows the fit $\sigma_{\rm y} \approx
(5400 \ \Phi^{1/2})$Pa. } \label{yield}
\end{figure}
%%%%

%%%%
\begin{figure} %[h]
\centerline{\resizebox{0.45\textwidth}{!}{
\includegraphics{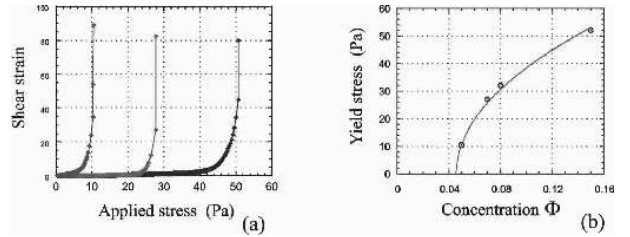} }  }
 \caption{The yield stress effect: (a) Strain-stress curves at a frequency
$\omega =1$~Hz for the MBBA colloid at $\Delta T \approx 15^{\rm
o}$ below $T_{\rm ni}$, for particle concentrations $\Phi=5\%, \,
7\%$ and $15\%$.  \ (b) The yield stress vs particle
concentration; solid line shows the fit $\sigma_{\rm y} \approx
(168 \ [\Phi-0.045]^{1/2})$Pa. } \label{yield2}
\end{figure}
%%%%

A different type of rheological experiment, a stress sweep at
constant low frequency $\omega=1$~Hz, allows to investigate the
yield stress behaviour. In figs.~\ref{yield}(a) and
\ref{yield2}(a) we plot the values of shear strain in response to
the gradually increasing stress between the rheometer plates. The
slope of the initial linear increase gives the value of $G'$.
However, at higher stress the material yields. We identify the
corresponding value of stress as the yield-stress $\sigma_{\rm
y}$, which is a function of colloid concentration. Once again, one
finds a difference of more than an order of magnitude between the
5CB and MBBA materials.

Plotting the concentration dependence $\sigma_{\rm y}(\Phi)$,
figs.~\ref{yield}(b) and \ref{yield2}(b), we again find that the
MBBA colloid shows a concentration threshold at $\Phi \sim 4.5\%$
[compare with fig.~\ref{modulus2}(b)]. A good single-parameter fit
to a square-root dependence $\sigma_{\rm y}\propto \Phi^{1/2}$ is
different from the scaling predicted by a crude theoretical model
in Appendix. However, the magnitude of yield stress is in the
correct range, appropriately much greater for 5CB colloids where
the strong nematic field should generate a greater pressure on
interfaces and thus preserve the particle dense packing.

\section{Elastic response of cellular superstructure}\label{elsec}

 First of all, one needs to appreciate the very large
elastic modulus, which usually tells about a significant energy
density stored in the aggregated nematic colloid. To illustrate
the values, let us compare the characteristic energy scales
involved. Assume, for the purpose of estimate,  a 5\% volume
fraction of colloid particles with $150 \, \hbox{nm}$ radius. The
director field around each particle contains the elastic energy
$\sim 10 \, KR$ in the case of strong anchoring, or $\sim 0.2 \,
W^2R^3/K$ in the case of weak anchoring, \cite{ukra}. Therefore,
even the energetically most unfavourable case of separate,
equidistant particles (separation $d \sim 4R$ at 5\%) results in
the energy density $\sim 0.16 \, K/R^2 \sim 60 \, \hbox{Pa}$. In
reality, particles would always aggregate in some way,
significantly reducing this estimate (e.g. particles lumped in the
nodes of disclination network in \cite{martin} are separated by $d
\sim 100 \, \mu\hbox{m}$, giving the energy density of just $\sim
0.01 \, \hbox{Pa}$). Also, considering that the particles of this
size will have the colloid parameter $WR/K \ll 1$, it is far more
likely that the case of weak anchoring would occur and even the
separate equidistantly distributed colloid would have the elastic
energy density of only $\sim 3\times 10^{-3} W^2/K \sim 10^{-4} \,
\hbox{Pa}$. None of these values can even crudely account for the
large values of elastic modulus $G'$ observed in nematic colloids.

Of course, it is not the energy density itself, but the rate of
its change -- a slope of energy variation with deformation, that
determines the modulus. We now try to outline the physical origin
of such variation. It seems clear that the dramatic enhancement of
mechanical properties is due to the cellular structure of
phase-separated nematic colloids. Particles, densely packed and
solidified on thin interfaces, produce very high surface tension
$\gamma$. There are several processes contributing to this surface
tension.

\begin{figure}%[b]
\centerline{\resizebox{0.45\textwidth}{!}{
\includegraphics{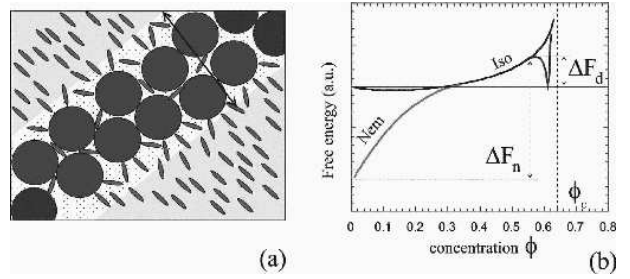}}}
\caption{(a) Model of cell border close-packed with colloid
particles -- the thin interface between clean nematic domains. The
interface thickness is related to the particle size, $d=nR$, with
$n\sim 4$ in the sketch. The remaining space in the interface is
filled with the matrix liquid in its isotropic state. \ (b) The
plot of free energy $F(\phi)$ (see \cite{no1} for detail) showing
the difference in local nematic mean field energy density $\Delta
F_{\rm n}$. The plot also shows the cusped minimum of depth
$\Delta F_{\rm d}$ resulting from the depletion locking between
particles in contact (see text and Appendix)} \label{wall}
\end{figure}

\noindent (1) Let us first consider the local melting of the
nematic phase. The gaps between closely packed particles are of
order or less than $\xi \sim 10 \, \hbox{nm}$, the characteristic
nematic correlation length. The nematic liquid in these small gaps
is melted, cf. fig.~\ref{wall}(a) and the phase diagram in
ref.~\cite{no1}. The thermodynamic energy density of a melted
nematic at $T<T_{\rm ni}$, $\Delta F_{\rm n}$ is obtained from the
mean-field energy part of eq.~(7) of \cite{no1}; qualitatively
\begin{equation}
\Delta F_{\rm n} \simeq \frac{(A_{\rm o} {T^*})^2}{4C} \,
\frac{1-\tau}{\tau} \sim 10^7 (1-\tau) \ \hbox{J/m}^3 ,
\label{dFn} \end{equation}
 with $\tau=T/T^*$. The last estimate in (\ref{dFn})
is based on the Landau parameters given in \cite{no1}. The
$(1-T/T^*)$ factor, of course, is only valid near the nematic
transition, where the variation of order parameter $Q^*(T)$ is
significant. The effective surface tension of such an interface is
a result of increasing the relative volume occupied by the dense
phase (assuming the constant interface thickness $nR$) with the
melted nematic in the gaps between particles:
\begin{equation}
\gamma_{\rm n} \approx (1-\phi_c) (nR) \Delta F_{\rm n},
\label{gn}
\end{equation}
where $(1-\phi_c) $ measures the amount of volume left for the
nematic liquid in the cell wall ($\phi_c$ being the particle
close-packing volume fraction).

\noindent (2) The imbalance in energy density between the pure
nematic inside cells on both sides of isotropic interfaces and the
isotropic melt inside these densely packed interfaces acts as an
effective pressure of a magnitude $\sim \Delta F_{\rm n}$. This
effect would compress the interfaces, trying to reduce the volume
occupied by the unfavourable melted state. This situation is
analogous to the process of packing of granular matter by gravity
or external pressure. Taking our monodisperse spherical particles,
the results are that the maximal particle volume fraction is
reached at random close packing $\phi_c \approx 0.64$, when there
are on average the maximal possible number of contacts, $N_c
\approx 8.6$, between the spheres. The contacts create a depletion
gap in the free energy, see Appendix and fig.~\ref{wall}(b). The
resulting cusped minimum is very narrow because the dependence of
packing concentration on the number of contacts is very sharp
\cite{torquato,davies}. These, and other theories of granular
matter provide a number of advanced estimates, some valid over a
large range of concentrations well below $\phi_c$. One may take a
view that a crude linear estimate $N_c \approx N_{\rm max}+\beta
(\phi-\phi_c)$ can be obtained simply by considering that $N_c
\rightarrow 0$ already at $\phi \sim 0.5$. This gives the (lower
bound) estimate for the slope: $\beta \sim 100$.

%%%%
%%%%

 The narrow cusped structure of the free energy
at the high-$\phi$ end of phase diagram provides the additional
reason for the effective surface tension of interfaces,
$\gamma_{\rm d}$, and also gives the expression for the yield
stress $\sigma_{\rm y}$. The interface tension is estimated by the
following argument: Take the value for the energy of a single
depletion gap eq.~(\ref{depletion}) and use the estimated linear
dependence for the number of contacts $N_c(\phi)$. We can now
describe the local free energy density increase in response to the
change (decrease) in local particle volume fraction
$(\phi-\phi_c)$
\begin{eqnarray}
F \approx  \frac{(\gamma_{\rm o} R \, a)\phi_c}{v_R}\, N_c =F_{\rm
min} +\frac{ \beta (\gamma_{\rm o} R \, a)\phi_c}{v_R}
(\phi_c-\phi) . \label{dFd}
\end{eqnarray}
The particle concentration can be related to the local change in
surface area $\delta {\cal A}$, namely $\phi-\phi_c \approx
-\phi_c \delta {\cal A}/{\cal A}_c$, assuming its constant
thickness and the ``equilibrium'' close-packing area ${\cal
A}_c$\footnote{Take the current density of interface of constant
thickness $nR$ to be $\phi=N/V$. Then $$\phi = \frac{N}{nR ({\cal
A}_c+ \delta {\cal A})} \approx \frac{N}{nR \, {\cal
A}_c}\left(1-\frac{\delta {\cal A}}{{\cal A}_c} \right) \equiv
\phi_c (1-\delta {\cal A}/{\cal A}_c).$$}. Then the change in the
full free energy on stretching the interface can be written as
\begin{eqnarray}
&& (F-F_{\rm min}) V  \approx \frac{ \beta \gamma_{\rm o} R \, a
\phi_c^2}{v_R {\cal A}_c}V  \, \delta {\cal A} \sim \frac{ \beta
\gamma_{\rm o} a \phi_c^2 \lambda }{R^2} \, \delta {\cal A}
\label{tense} \\ && \gamma_{\rm d} \approx \frac{\beta \gamma_{\rm
o} a \phi_c^2 \lambda }{R^2}  \nonumber
\end{eqnarray}
where $v_R$ is the particle volume and $\lambda = V/{\cal A}_c$
the cell size. In this way, the eq.~(\ref{tense}) identifies the
effective surface tension $\gamma_{\rm d} $ resulting from a local
energy increase due to the rapid loss of particle contacts on
stretching the interface.

We have a cellular solid system with interfaces under tension
(rather then, say, elastic walls as in \cite{ashby}), with open
cells, so that the conservation of cell volume is not posing a
constraint. The elastic modulus of such a system is estimated as
$G' \simeq \gamma/\lambda$, where $\gamma=\gamma_{\rm
n}+\gamma_{\rm d}$ is the total effective surface tension of cell
walls and the mesh size $\lambda \sim nR/\Phi$. We thus obtain
from the two physically different contributions to the surface
tension:
\begin{eqnarray}
G' &\approx & (1-\phi_c)\frac{(A_{\rm o} {T^*})^2(1-\tau)}{4C
\tau}  \, \Phi + \frac{ \beta \gamma_{\rm o} a \phi_c^2}{R^2}
\label{Gmodulus} \\ &\sim & \ [10^6 (1-\tau)\, \Phi + 10^5] \,
\hbox{Pa} \ . \nonumber
\end{eqnarray}
The numerical estimate is obtained for the material parameters of
5CB, cf. the Appendix of ref.~\cite{no1}. The experimental data of
figs.~\ref{modulus}(b) and \ref{modulus2}(b) suggest that deep
below the nematic transition of a classical thermotropic nematic
material, the dominant factor is the energy of nematic melting
and, thus, the elastic modulus $G'$ is a linear function of
colloid concentration $\phi$. On the other hand, close to $T_{\rm
ni}$, when $\tau \rightarrow 1$, the modulus should become a
concentration-independent material constant determined by the
packing of hard spheres under pressure.

 One of the consequences of the depletion mechanism is the effect
of the nematic order $Q(T)$, and of small vibrations. This
mechanism is based on the particle packing approaching the random
close packing limit $\phi_c$ and having a very sharp dependence of
average number of depletion contacts on local concentration. It is
known that the packing density increases sharply when the sandpile
is subjected to an external pressure (the role often played by the
gravity body force) and the vibrations. In our case, the nematic
mean field creates such an effective pressure. We expect, and
indeed find in the experimental study of cellular structure
\cite{no1}, that a small-amplitude oscillating strain in the
rheometer makes the cells much more uniform and interfaces visibly
sharper.

\section{Conclusions}  \label{concl}
We have reported results of investigation of mechanical properties
of new cellular solid aggregates of liquid crystal colloids
quenched below their clearing point. Two main conclusions can be
made from this study. The magnitude of the low-frequency storage
modulus, which is comparable with that of a normal rubber,
indicates that the thin cell walls possess an unusually high
effective surface tension. Theoretical arguments suggest that the
main mechanism for such a local increase in energy density is the
local melting of nematic liquid, making $G'$ a linear function of
average colloid concentration. The second important result of this
paper is the yield behaviour of cellular macrostructure and the
dependence of the yield stress on concentration. Qualitative
arguments suggest a possible mechanism for the observed yield
stress. However, within the limited accuracy of our experiments,
we observed a concentration dependence $\sigma_{\rm y} \propto
\Phi^{1/2}$ which is not accounted for by the crude depletion
model.

An interesting comparison can be made between the colloids based
on ``good'' and ``poor'' nematic matrix. Not only the values of
elastic modulus and the yield stress are much lower in the MBBA
system, but we also observe a characteristic threshold
concentration $\Phi \sim 4.5\%$ below which the MBBA colloid does
not phase separate into the cellular structure.

We must emphasize a number of experimental difficulties in the
study of mechanical properties of our cellular solids. The
rheometric results, even for the same batch of colloid, are often
not reproducible. The reasons are the important role of the rate
of quenching and, specifically for the parallel-plate rheometric
experiment, slipping of plates lubricated by the nematic liquid.
It is possible that an alternative approach, for instance
examining the mechanical response to an indentation probe rather
than the sheared plate, will give a better defined experiment.

Notwithstanding these uncertainties, we have observed the unusual
phase behaviour and remarkable rheological response in the
``classical'' liquid crystal colloid -- a straightforward
low-concentration mixture of thermotropic nematic with sterically
stabilised PMMA particles. The metastable cellular solid
aggregates are preserved by thermodynamic forces provided by the
nematic mean field. No doubt, further investigations into the
mechanisms of cellular structure formation and its stability would
make possible a controlled, reliable and robust preparation of a
wide variety of new materials. \\

\noindent We appreciate valuable discussions with M.E. Cates, S.M.
Clarke and P.D. Olmsted.  This research has been supported by
EPSRC UK.

\appendix
\section*{Appendix:\\ Depletion gap in densely packed walls}
 Another effect that switches on when the particles come into
close contact on localised interfaces is the depletion reduction
of their surface energy. When isolated in the nematic solvent
matrix, each particle's surface has the energy of the order
$\gamma_{\rm o} (4\pi R^2)$, with $\gamma_{\rm o}$ the main
surface tension of, e.g. hydrocarbon PHSE and biphenyl oil
interface; qualitatively the value of this tension is of the order
$\gamma_{\rm o} \sim 0.05 \, \hbox{J/m}^2$ \cite{polybase} and far
exceeds all anisotropic anchoring corrections. In the Deryagin
approximation, each contact between two particles creates a
depletion disk and reduces this energy by $\sim 2 \gamma_{\rm o} R
\, a$, where $a \sim 1 \, \hbox{nm}$ is the size of the molecule
of suspending liquid. In the randomly close-packed state each
particle has $N_c \sim 8.6$ contacts (the never-reached absolute
maximum would be $N_c=12$ for fcc packing) which, for a cell wall
of thickness $nR$, gives the depletion gain in the local potential
energy density
\begin{equation}
\Delta F_{\rm d} \sim N_c(\gamma_{\rm o} R \, a)\frac{\phi_c}{R^3}
\ \sim  5\times 10^{3} \hbox{J/m}^3 \label{depletion}
\end{equation}
This gap would continuously deepen on increasing the number of
contacts. However, the maximal number $N_c$ is reached at close
packing and cuts this linear decrease, thus creating a cusped
minimum of the free energy density, shown in fig.~\ref{wall}(b).
The depth of the minimum (\ref{depletion}) gives an estimate for
the yield stress: when the interface is distorted such that the
particles lose most of their contacts, the high resistance to
deformation is lost and the cellular structure would no longer be
stable. Taken in proportion to the total amount of interface in
the cellular structure, the macroscopic yield stress should be a
function of initial concentration of the colloid: $$ \sigma_{\rm
y} \sim \Phi \, \Delta F_{\rm d} \ \sim 200 \, \hbox{Pa}.$$ One
should appreciate that the estimates here are extremely crude, and
yet they give not an unreasonable value $\sigma_{\rm y} \ll G'$.

\end{document}